\documentclass[twocolumn,showpacs,preprintnumbers,amsmath,amssymb,prl]{revtex4}
\usepackage{graphicx}     
\usepackage{dcolumn}      
\usepackage{bm}           
\usepackage{color}        
\def\sb#1{$_{#1}$}
\def\sp#1{$^{#1}$}

\def\cete{CeTe$_{3}$}
\begin{document}
%
%
\title{%
Local atomic structure and discommensurations in the charge density
wave of {CeTe$_{\boldmath{3}}$} }
\author{H.~J. Kim$^1$, C.~D. Malliakas$^2$,
A. Tomic$^1$, S.~H. Tessmer$^1$, M.~G. Kanatzidis$^2$ \& S.~J.~L.
Billinge$^{1*}$ }
\affiliation{
$^1$Department of Physics and Astronomy, Michigan State University, 
East Lansing, Michigan 48824-1116, USA\\
$^2$Department of Chemistry, Michigan State University, 
East Lansing, MI~~48824, USA}%
\date{\today}
\begin{abstract}
The local structure of \cete\ in the incommensurate charge density
wave (IC-CDW) state has been obtained using atomic pair distribution
function (PDF) analysis of x-ray diffraction data. Local atomic
distortions in the Te-nets due to the CDW are larger than observed
crystallographically, resulting in distinct short and long Te-Te
bonds.  Observation of different distortion amplitudes in the local
and average structures are explained by the discommensurated nature
of the CDW since the PDF is sensitive to the local displacements
within the commensurate regions whereas the crystallographic result
averages over many discommensurated domains.  The result is
supported by STM data. This is the first quantitative local
structural study within the commensurate domains in an IC-CDW
system.
\end{abstract}
\maketitle

Incommensurate charge density waves (IC-CDWs) are a fundamental
property of low-dimensional metals~\cite{grune;b;dwis94} and also
underly the novel properties of correlated electron oxides such as
cuprates in the pseudo-gap
state~\cite{versh;s04,hoffm;s02,hanag;n04}, and manganites at high
doping~\cite{loudo;prl05}. Knowing the nature of local atomic
displacements (Peierls distortions) in the IC-CDWs is crucial to
understand such factors as electron-lattice
coupling~\cite{milwa;n05}, yet this information is difficult to
obtain quantitatively. Here we solve this problem by taking the
novel approach of using a local structural method, the atomic pair
distribution function (PDF) technique~\cite{egami;b;utbp03}, to
determine the local atomic displacements with high precision in the
system \cete. IC-CDWs, and the underlying atomic displacements, can
be uniform incommensurate modulations or locally commensurate waves
separated by narrow domain walls, known as
discommensurations~\cite{mcmil;prb76}, where the phase of the wave
changes rapidly. Here we show that the IC-CDW in \cete\ is
discommensurated and obtain  for the first time the quantitative
local atomic displacements within the commensurate domains.  

In the case of \emph{incommensurate} CDWs, superlattice peaks
observed crystallographically yield the average distorted structure.
Except in the cases where the domains are periodically arranged,
giving rise to satellite peaks~\cite{monct;prb75}, it is not
possible to determine whether the underlying CDW is truly
incommensurate or forms a discommensurated structure with
commensurate regions separated by domain walls~\cite{mcmil;prb76}. A
number of techniques have been successful at differentiating between
the truly incommensurate and discommensurated cases. The earliest
verification of a discommensurated phase came from photoemission
spectroscopy evidence that the Ta 4$f$ states in 1$T$-TaS$_2$ had
the same splitting in the commensurate and nearly-commensurate
states~\cite{hughe;cop76}. Photoemission is a local probe and found
distinct Ta environments rather than a broad continuum expected from
a purely incommensurate state. Similarly, another local probe,
nuclear magnetic resonance (NMR), found distinct Knight-shifts for
three Se sites in the incommensurate state of 2$H$-TaSe$_2$, similar
to the commensurate phase~\cite{suits;prl80,suits;prb81}. High
resolution atomic imaging methods have also contributed to this
debate. The strain fields due to the domain walls were observed in
dark field transmission electron microscopy (TEM)
measurements~\cite{chen;prl81}. Interestingly, atomic resolution
images in real-space have difficulty in resolving discommensurated
domains~\cite{gibso;prl83,ishig;prb91,kuwab;pssa86,steed;u86}.
However, Fourier analysis of scanning tunneling microscopy (STM)
images can be a reliable measure, as discussed in detail by
Thomson~$et~al.$~\cite{thoms;prb94}.

As in the case of the NMR and photoemission studies, the PDF
approach described here makes use of the fact that the local
structure deviates from the average in the discommensurated case. By
comparing atomic displacements determined from the PDF with those
determined crystallographically we establish the presence of
commensurate domains, but crucially, also obtain quantitatively  the
atomic structure within these domains.  This novel approach is here
applied to the incommensurate phase of CeTe$_{3}$.

\begin{figure}
\includegraphics[width=2.7in,keepaspectratio=1]{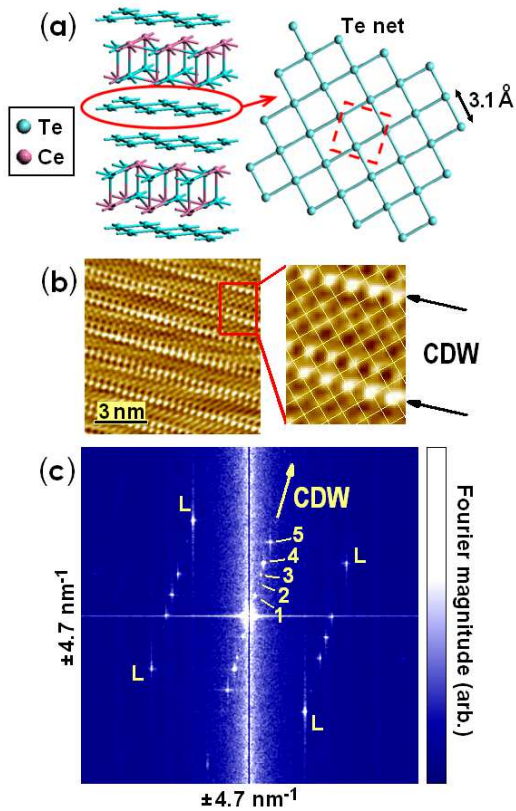}
\caption{\label{fig;stm} (a) The crystal structure of CeTe$_{3}$
with the square Te net that supports the CDW highlighted. The
reduced unit cell on the Te net is indicated by the red dashed box
(\textbf {\textcolor{red}{- -}}). (b) A representative STM image
from the square Te net showing the CDW.  On the expanded image, the
network of Te bonds is superimposed.  (c) The Fourier transform of
the STM data. To achieve a high signal-to-noise ratio, the transform
represents the average of 24 images (each image was
$27~{\times}~27$~nm).  The square Te net gives rise to four distinct
peaks (L), with peaks related to the CDW oriented at $45^\circ$, as
indicated by the arrow.  The fundamental CDW peak (corresponding to
a wavelength of $\approx 15$~\AA) and the $\lambda/2$ harmonic are
labeled 1 and 3, respectively. Peaks 2 and 4 are in close proximity
to 3, implying a characteristic discommensuration length of 38~\AA,
as described in the text.  Peak 5 corresponds to the diagonal of the
Te net.  This component of the Te net may be enhanced due to the
underlying crystal structure; the CDW-lattice interaction may also
enhance this peak.}
\end{figure}

In its undistorted form, CeTe$_{3}$ takes the NdTe$_{3}$ structure
type with space group $Cmcm$~\cite{lin;ic65}. It forms a layered
structure with ionic [Ce\sb{2}\sp{3+}Te\sb{2}\sp{2-}]\sp{2+} layers
sandwiched between two Te\sp{-} layers. These sandwich layers stack
together with weak van der Waals forces to form the 3-dimensional
structure. Te ions in the Te\sp{-} layers form a square-net with
3.1~\AA\ Te-Te bonds. The structure is shown in
Fig.~\ref{fig;stm}(a).
The electronic bands crossing the Fermi level are Te $p$-bands from the
2D square nets~\cite{dimas;prb95} and the CDW forms in these
metallic layers. In the CDW state an incommensurate superlattice is
observed~\cite{malli;jacs05}, with a wavevector characteristic of a
strong Fermi-surface nesting vector in the electronic
structure~\cite{gweon;prl98,broue;prl04,komod;prb04,laver;prb05}.
This is a surprisingly stable and simple single-$q$ IC-CDW state in
an easily cleavable 2D square net making the  RETe$_3$ (RE=Rare
Earth) systems ideal for studying the IC-CDW
state~\cite{dimas;prb95}. The atomic distortions giving rise to the
superlattice have been solved crystallographically from single
crystal x-ray diffraction data~\cite{malli;jacs05}. The
incommensurate wavelength of the distortion is close to $25a/7$,
where $a$ is the lattice parameter of the undistorted phase. The
distorted structure is in the $Ama2$ spacegroup~\cite{malli;jacs05}.
From the crystallography alone it is not possible to determine
whether this distorted structure is truly incommensurate or whether
discommensurations form between short-range commensurate domains.

The X-ray PDF experiment was conducted on a fine powder of \cete\
prepared as described in Ref.~\cite{malli;jacs05}. CeTe$_{3}$ powder
 was loosely packed in a flat plate with thickness of 1.0~mm sealed
with kapton tape.  Care must be taken when grinding this material or
turbostratic disorder is introduced, significantly modifying the
stacking of the layers.  Diffraction data were collected at 300~K
using the rapid acquisition pair distribution
function (RA-PDF) technique~\cite{chupa;jac03}.  
Standard corrections~\cite{chupa;jac03,egami;b;utbp03} were made
using the program PDFgetX2~\cite{qiu;jac04} to obtain the properly
normalized total scattering function, $S(Q)$,~\cite{egami;b;utbp03}
which was truncated at $Q_{max}$ of 25~\AA$^{-1}$ before Fourier
transforming to obtain the PDF, $G(r)=
\frac{2}{\pi}\int_{0}^{\infty} Q [S(Q)-1] \sin (Qr)\> dQ$.
Structural models are fit to the data using the program
PDFFIT~\cite{proff;jac99}.

\begin{figure}
\includegraphics[width=2.4in,keepaspectratio=1]{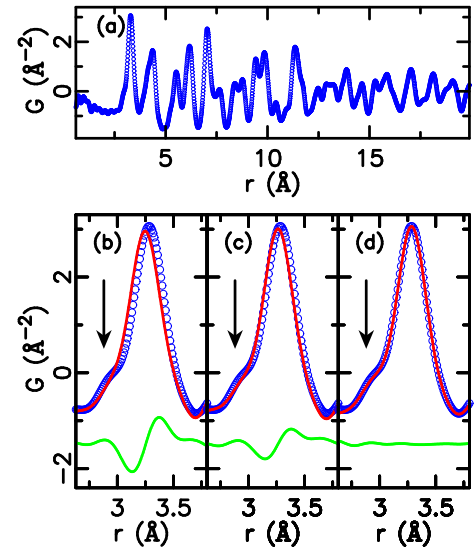}
\caption{(a) The PDF of CeTe$_{3}$ at room temperature.  In (b)-(d)
the first peak of the experimental PDF of CeTe$_{3}$ (\textbf
{\textcolor{blue}{$\circ$}}) is plotted with the calculated PDF
(\textbf {\textcolor{red}{--}}) from various models: (b) the
undistorted crystal structure model ($Cmcm$), (c) the distorted
crystallographic model and (d) the local structural model determined
by the PDF refinement over the range $2.5 < r < 6.37$~\AA.  The
difference between the experimental and calculated PDFs (\textbf
{\textcolor{green}{--}}) is plotted below the data in each panel.
The shoulder due to the Peierls distortions in the Te-nets is
indicated by an arrow. } \label{fig;first PDF peak}
\end{figure}

The PDF of \cete , measured at room temperature, is shown in
Fig.~\ref{fig;first PDF peak}(a).
The PDF gives the probability of finding an atom at a
distance-\emph{r} away from another atom. The nearest neighbor peak
around 3.1~\AA\ comes from the Te-Te bond in the nets and the Ce-Te
bond in the intergrowth layers. This is shown on an expanded scale
in Figs.~\ref{fig;first PDF peak}(b)-(d). A shoulder is evident on
the low-$r$ side of the peak. This feature is robust; it is much
larger than the statistical and systematic errors and is reproduced
in measurements of isostructural compounds NdTe\sb{3} and
PrTe\sb{3}. Fig.~\ref{fig;first PDF peak}(b) shows the fit to this
peak of the undistorted crystal structure model ($Cmcm$), where only
symmetry allowed atomic positions and isotropic thermal factors were
refined. The result clearly does not explain this shoulder which
originates from short Te-Te bonds in the Te-net. Surprisingly
however, the PDF calculated from the \emph{distorted} structure
determined crystallographically~\cite{malli;jacs05} also does not
explain this shoulder well.  In this case the atoms were fixed at
the crystallographically determined positions and isotropic thermal
factors were refined. This resulted in a better fit to the first
peak (Fig.~\ref{fig;first PDF peak}(c)); however, the fit is not
ideal and required a large value of $U_{iso}$ for the Te atoms in
the nets ($U_{iso}=0.0152(2)$~\AA$^{2}$). The value was two times
larger than $U_{iso}$ of the Ce and Te atoms in the ionic
[Ce\sb{2}\sp{3+}Te\sb{2}\sp{2-}]\sp{2+} layers (0.0077(2)~\AA$^{2}$
and 0.0080(2)~\AA$^{2}$ for $U_{iso}$ of Ce and Te atoms,
respectively).

The large fluctuation in the difference curve in Fig.~\ref{fig;first
PDF peak}(b) arises because the real distribution of Te-Te
bond-lengths in the data is broader than in the undistorted model.
This fluctuation in the difference curve is smaller in
Fig.~\ref{fig;first PDF peak}(c) because the distortions of the Te
net in the $Ama2$ crystallographic model result in a broader Te-Te
bond-length distribution. However, clearly the
distorted-crystallographic model still has a Te-Te bond-length
distribution that is narrower than in the data. We therefore refined
the Te-net distortions directly in the PDF by allowing the atomic
positions in the model to vary.  The model was constrained to have
the $Ama2$ symmetry and the same unit cell was used as in the
distorted crystallographic model.   The refinement result for $2.5 <
r < 6.37$~\AA\ is shown in Fig.~\ref{fig;first PDF peak}(d).  As
well as giving a significantly better fit to the low-$r$ region of
the PDF, this refinement resulted in much smaller and more physical
thermal factors on the planar Te ions.

The model of the local structure refined from the PDF gives a
broader range of Te-Te bond lengths (from 2.83~\AA\ to 3.36~\AA )
than the crystallographic distorted model (from 2.94~\AA\ to
3.26~\AA ).  It is also interesting to see the shape of the
bond-length \emph{distributions} for the Te-Te bonds in the Te-nets
from these two models. These are shown in Fig.~\ref{fig;bond
distributions of Te-nets}.
\begin{figure}
\includegraphics[width=2.4in, keepaspectratio=1]{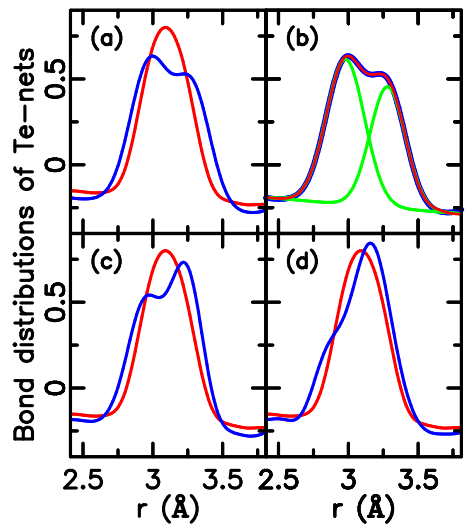}
\caption{Bond-length distributions in the Te-nets refined from the
PDF over various $r$-ranges (\textbf {\textcolor{blue}{--}}) (a)
$r_{max}$=6.37~{\AA} (b) $r_{max}$=6.37~{\AA} (c)
$r_{max}$=14.5~{\AA} (d) $r_{max}$=27.1~{\AA}.  The $r_{min}$ value
was fixed to 2.5~{\AA} for all the cases.  The bond distribution
from the distorted crystallographic model (\textbf
{\textcolor{red}{--}}) is plotted in (a), (c), and (d) for
comparison.  In (b) the bond distribution of the local structural
model ($r_{max}$=6.37~{\AA}) is fit with two Gaussians.  The fit is
shown as a red line (\textbf {\textcolor{red}{--}}) and the two
Gaussian sub-components in green (\textbf {\textcolor{green}{--}}).
} \label{fig;bond distributions of Te-nets}
\end{figure}
The blue solid line shows the bond-length distribution refined from
the PDF and the red line is the distribution from the
crystallographic model~\cite{malli;jacs05}.  For direct comparison
the distributions are plotted using the same thermal broadening of
0.007~\AA$^{2}$. The distorted-crystallographic model has broad, but
symmetric and Gaussian bond-length distribution coming from the
continuous distribution of Te-Te bond lengths in the \emph{average}
IC-CDW. On the other hand, the local structure refinement
($r_{max}=6.37$~\AA ) yields a bond-length distribution that is
clearly bimodal and is separated into distinct ``short" and ``long"
Te-Te distances.  This is emphasized in Fig.~\ref{fig;bond
distributions of Te-nets}(b) where we show a fit of two, well
separated, Gaussian curves to the PDF-refined bond-length
distribution.  This behavior is characteristic of oligomerization
with Te forming bonded and non-bonded interactions with its
neighbors in the net~\cite{patsc;pccp02} that would be expected in a
commensurate structure.  Since we know that the modulation is
incommensurate on average, this is strong evidence that the
structure consists of commensurate domains separated by
discommensurations. As $r_{max}$ in the PDF-refinements is increased
the PDF refined distribution crosses over towards the
crystallographic result and by $r_{max}=27.1$~\AA , resembles it
rather closely (Fig.~\ref{fig;bond distributions of Te-nets}(d)).

We have applied STM on the exposed Te net of a cleaved single
crystal of \cete,  grown according to the method described in
Ref.~\cite{malli;jacs05}.  Measurements were done at 300~K in the
constant current mode of the STM. Data were acquired with a bias
voltage of 100 mV and with a tunneling current of 0.6 nA.  
Fig.~\ref{fig;stm}(b) shows a representative atomic resolution image
with the CDW modulation clearly visible oriented at 45$^\circ$ to
the net.  To investigate the images for discommensurations, we examine the
corresponding two-dimensional Fourier transform, shown in
Fig.~\ref{fig;stm}(c).  As indicated by the labels, in addition to
the fundamental CDW peak (1), four more peaks lie along the CDW
direction (2-5).  Although the transforms of real space images
resemble diffraction data, symmetry requirements intrinsic to
diffraction data do not apply.  As demonstrated by Thomson and
co-workers, the Fourier transforms of STM images that exhibit true
discommensurations always have extra peaks in proximity to the
fundamental CDW peak~\cite{thoms;prb94}.  This arises from the fact
that a discommensurate CDW can be expressed as the product of a
uniformly incommensurate CDW and a modulation
envelope~\cite{thoms;prb94}.  The wavelengths of the envelope are
given by the differences in the wave vectors of closely spaced
peaks.  The longest such wavelength in our images is 38~\AA,
corresponding to peaks 2-3 and 3-4 in the Fourier transform,
indicating that a discommensuration separation of this length-scale
exists.  This is consistent with the refined PDF behavior which
crosses over from the local to the average behavior for a refinement
range of around 27~\AA (Fig.~\ref{fig;bond distributions of
Te-nets}), which would be expected to occur at around, or a little
above, the \emph{radius} of the local domains.

As well as the bond-length distributions, the local and average
structure refinements allow us to study the patterns of atomic
displacents due to the IC-CDW.
\begin{figure}
\centering
\includegraphics[width=2.4in,keepaspectratio=1]{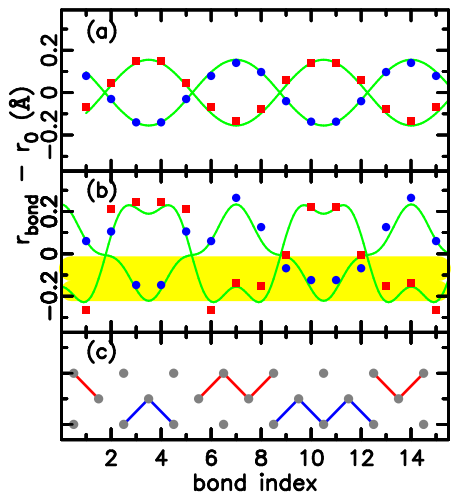}
\caption{Te-Te bond length deviation from the average value as
refined crystallographically (a) and from the PDF (b). The deviation
$r_{bond}-r_0$ is defined such that $r_{i}$ is the Te-Te bond length
of the $i^{th}$ bond (bond-index $i$) in the unit cell and
$r_0=3.1$~\AA . (c) Schematic of the arrangements of ``short" bonds
within the unit-cell highlighting the formation of oligomers. Short
bonds are defined as those lying within $\pm$~$\sigma$ of the center
of the first Gaussian of the bimodal distribution in
Fig.~\ref{fig;bond distributions of Te-nets}~(b), indicated as a
yellow band in the Figure. The Gray markers represent Te atoms. Red
(blue) lines and markers are bonds lying in the top (bottom) row of
the unit cell. The CDW is out of phase between these two rows of
bonds.} \label{fig;spatial bond length distribution}
\end{figure}
The average structure refinement~\cite{malli;jacs05} results in an
almost perfectly sinusoidal pattern of bond-lengths, with the
wavelength of the CDW (Fig.~\ref{fig;spatial bond length
distribution}(a)), clearly identifying these as Peierls distortions.
The local structural model was refined in the same unit cell and
space-group, but results in a much more square-wave like
distribution, consistent with the distinct short and long Te-Te
distances described above (Fig.~\ref{fig;spatial bond length
distribution}(b)). Fig.~\ref{fig;spatial bond length
distribution}(c) shows the pattern of bonded Te-Te atoms that
results when the short distances determined from the PDF data are
plotted in the unit cell. In this way, the Peierls distortions due
to the IC-CDW can be thought of as forming oligomers in the Te net.
In this picture, the discommensurations occur when the pattern of
oligomers has defects. This is a common picture in the chemistry
literature~\cite{patsc;jacs01,malli;jacs05}, though we note that
this picture is not supported by the crystallographic results shown
in (Fig.~\ref{fig;spatial bond length distribution}(a)) and needed
the application of a local structural method to show that it has a
physical reality beyond its heuristic value.  

The refined parameters of the low-$r_{max}$ PDF refinements yield
quantitatively the atomic displacements within the commensurate
domains. This is the first demonstration of
the use of the PDF to obtain quantitatively the atomic displacements
(Peierls distortion) within the commensurate domains of a
discommensurated IC-CDW. This opens the way to a quantitative first
principles calculations and a better microscopic understanding of
the IC-CDW state.

We gratefully acknowledge P.~M.~Duxbury, S.~D.~Mahanti, and
D.~I.~Bilc for discussions and D. Robinson and D. Wermeille for help
with collecting data.  Work was supported by the National Science
Foundation through grant DMR-0304391, DMR-0443785, and DMR-0305461.
MUCAT is supported by the US Department of Energy through contract
W-7405-Eng-82 and the APS by contract W-31-109-Eng-38.

\end{document}